 \documentclass[smallabstract,smallcaptions]{dccpaper}

\usepackage{epsfig}
\usepackage{amsmath}
\usepackage{amssymb}
\usepackage{color}
\usepackage{url}
\usepackage{multirow}
\usepackage{subfigure}
\newlength{\figurewidth}
\newlength{\smallfigurewidth}

\begin{document}

\title
{\large
\textbf{Learned Image Compression with Dual-Branch Encoder and Conditional Information Coding}
}

\author{%
Haisheng~Fu$^{ \ast \dag}$, Feng~Liang$^{\ast}$, Jie~Liang$^{\dag}$, Zhenman~Fang$^{\dag}$,\\[0.5em] Guohe~Zhang$^{\ast}$,  and Jingning~Han$^{\star}$\\[0.5em]
{\small\begin{minipage}{\linewidth}\begin{center}
\begin{tabular}{ccc}
$^{\ast}$Xi’an Jiaotong University &$^{\dag}$Simon Fraser University & $^{\star}$Google \\
\end{tabular}
\end{center}\end{minipage}}
}

\maketitle
\thispagestyle{empty}

\begin{abstract}
Recent advancements in deep learning-based image compression are notable. However, prevalent schemes that employ a serial context-adaptive entropy model to enhance rate-distortion (R-D) performance are markedly slow. Furthermore, the complexities of the encoding and decoding networks are substantially high, rendering them unsuitable for some practical applications. In this paper, we propose two techniques to balance the trade-off between complexity and performance. First, we introduce two branching coding networks to independently learn a low-resolution latent representation and a high-resolution latent representation of the input image, discriminatively representing the global and local information therein. Second, we utilize the high-resolution latent representation as conditional information for the low-resolution latent representation, furnishing it with global information, thus aiding in the reduction of redundancy between low-resolution information. We do not utilize any serial entropy models. Instead, we employ a parallel channel-wise auto-regressive entropy model for encoding and decoding low-resolution and high-resolution latent representations. Experiments demonstrate that our method is approximately twice as fast in both encoding and decoding compared to the parallelizable checkerboard context model, and it also achieves a 1.2\% improvement in R-D performance compared to state-of-the-art learned image compression schemes. Our method also outperforms classical image codecs including H.266/VVC-intra (4:4:4) and some recent learned methods in rate-distortion performance, as validated by both PSNR and MS-SSIM metrics on the Kodak dataset.
\end{abstract}

\Section{Introduction}

Image compression is a fundamental and crucial topic in the field of signal processing. Over the past few decades, several classical standards have emerged, including JPEG \cite{JPEG}, JPEG2000 \cite{JPEG2000}, BPG \cite{BPG}, and VVC, which generally follow the same transform coding paradigm: linear transformation, quantization, and entropy encoding.

Recent learned image compression methods \cite{GLLMM, He_2022_CVPR, He_2021_CVPR} have outperformed the current best classical image and video encoding standard VVC in terms of both peak signal-to-noise ratio (PSNR) and multi-scale structural similarity (MS-SSIM). This indicates that learned image compression methods hold tremendous potential for the next generation of image compression technologies.

Most learning-based image compression methods are Convolutional Neural Networks-based (CNN-based) approaches \cite{cheng2020, He_2022_CVPR, Asymmetric_Fu} that use the variational autoencoder (VAE) \cite{Variational}. However, with the recent development of vision Transformers \cite{vit}, several transformer-based learning methods \cite{zhu2022transformerbased, Zou_2022_CVPR, Liu_2023_CVPR} have been introduced. For instance, in CNN-based approaches, a residual block-based image compression model is proposed to achieve performance comparable to VVC in terms of PSNR. Meanwhile, in the realm of transformer-based methods, a swin-transformer-based image compression model has been proposed to enhance rate-distortion performance. Both CNN-based and transformer-based approaches offer distinct advantages: CNNs excel in local modeling with lower complexities, whereas transformers are adept at capturing non-local information. Nevertheless, the complexity of swin-transformer-based methods outperforms that of the CNN schemes.

The design of entropy models is also a critical aspect of learned image compression. A common approach introduces additional latent variables as hyper-priors, thereby transforming the compact encoded symbol probability model into a joint model \cite{Variational}. Based on this, several methods \cite{Joint, cheng2020, GLLMM} have been developed. For instance, masked convolution \cite{Joint} is proposed to capture context information. More accurate probabilistic models, such as GMM \cite{cheng2020} and GLLMM \cite{GLLMM}, have been introduced to enhance compression performance. Furthermore, a parallel channel autoregressive entropy model has been proposed \cite{channel}, wherein the latent representation is divided into 10 slices. The encoded slices can aid the encoding of subsequent slices by providing side information in a pipelined manner.

Recently, some attention modules \cite{chen2021, cheng2020} have been proposed to enhance image compression. Attention modules can be incorporated into the image compression framework to assist the model in focusing more on detailed information. However, many schemes are time-consuming or can only capture local information \cite{chen2021}. To reduce the complexity of the attention module, a simplified attention model is placed in the main encoder and decoder to enhance image compression. We also introduce an attention module \cite{cheng2020} to improve the rate-distortion performance.

The contributions of this paper can be summarized as follows:

First, we employ two branches of encoders to learn latent representations at different resolutions of the input. This strategy allows us to better capture both global and local information from the input image. Notably, each branch operates independently without any information exchange.

Second, we utilize the high-resolution latent representation as conditional information to assist in the encoding and decoding of the low-resolution latent representation. This strategy helps to remove spatial redundancy in the low-resolution latent representation, enhancing the overall performance of the framework. We do not utilize any serial entropy models. Instead, we employ a parallel channel-wise auto-regressive entropy model \cite{channel, Liu_2023_CVPR} for encoding and decoding low-resolution and high-resolution latent representations.

\begin{figure}[!t]
	\centering
		\includegraphics[scale=0.7]{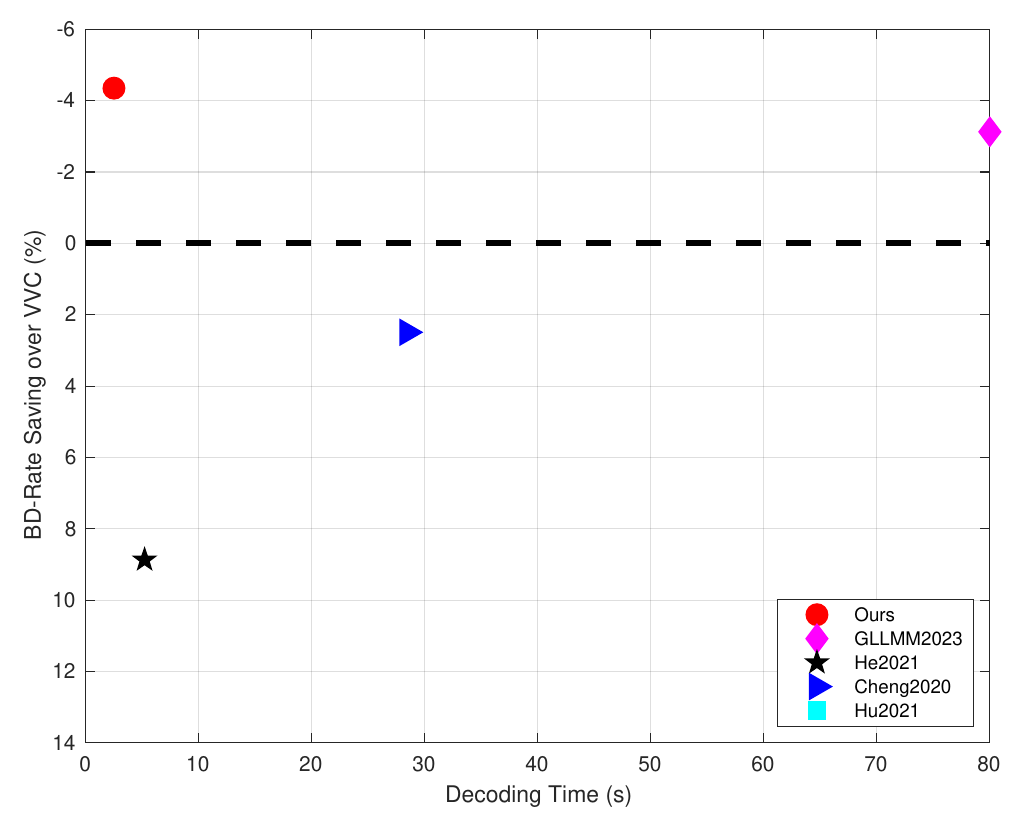}
	\caption{The decoding time and BD-Rate savings over H.266/VVC for different schemes are illustrated on Kodak dataset. A superior result is positioned in the upper-left corner. The notably extended decoding time for GLLMM \cite{GLLMM} is explicitly indicated in brackets.}
	\label{comp_time_bdrate}
\end{figure}

Third, extensive experiments demonstrate that our method achieves state-of-the-art performance when compared to recent learning-based methods and traditional image codecs on the Kodak dataset. Our method is approximately twice as fast in both encoding and decoding compared to the parallelizable checkerboard context model \cite{He_2021_CVPR}, and it also achieves a 1.2\% improvement in R-D performance compared to state-of-the-art learned image compression schemes. Our method also outperforms the latest classical image codec in H.266/VVC-Intra (4:4:4) and other leading learned schemes such as \cite{He_2021_CVPR} in both PSNR and MS-SSIM metrics. The decoding time and BD-Rate comparisons with VVC for various methods are presented in Fig. \ref{comp_time_bdrate}.

\section{The Proposed Image Compression Framework}

In this section, we describe the whole framework of the proposed method. Subsequently, we will detail its major components, including the dual-branch encoding network, conditional coding of low-resolution latent representations, and the associated training methodology.

\begin{figure*}[!thp]
	\centering
		\includegraphics[scale=0.45]{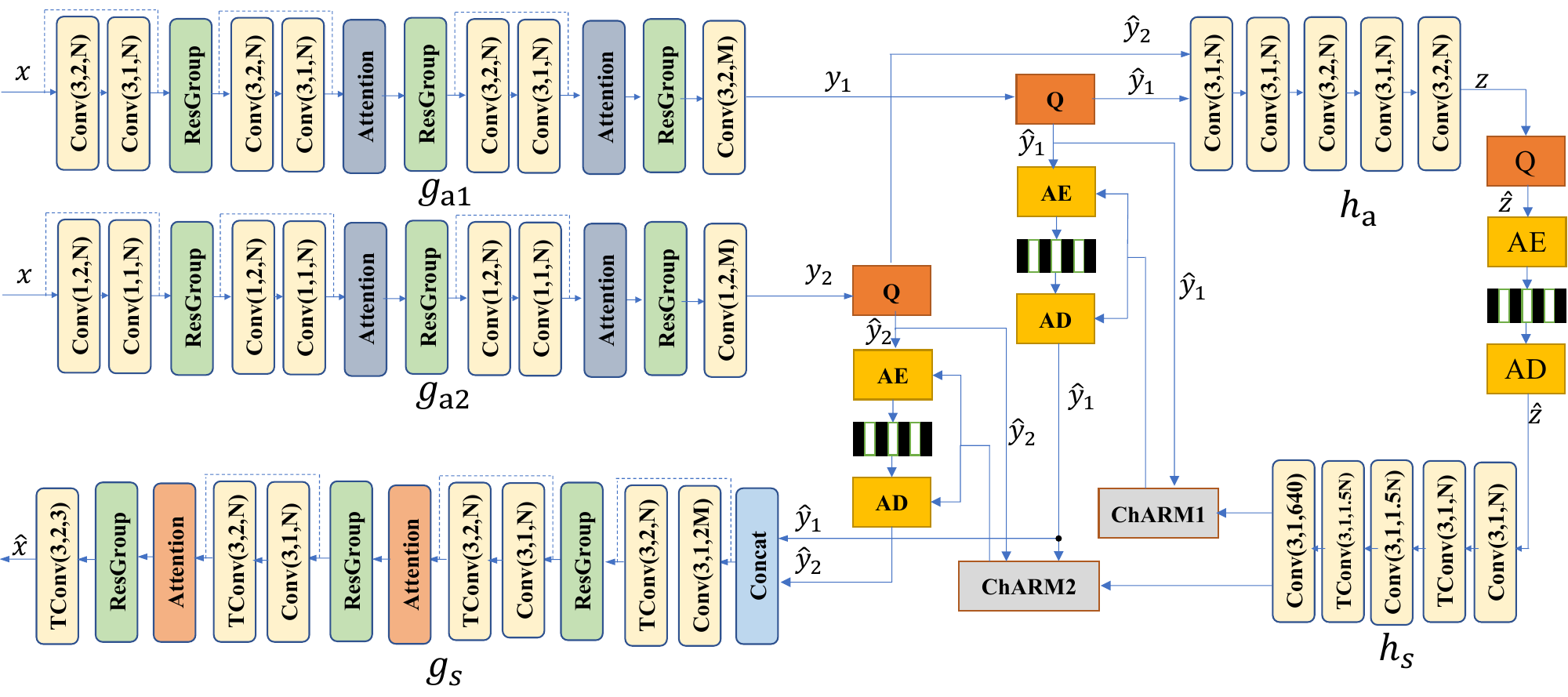}
	\caption{The overall architecture of our image compression framework. \textbf{ChARM} represents a
   channel-wise auto-regressive entropy model. \textbf{ResGroup} represents the residual group blocks.}
	\label{whole_networkstructure}
\end{figure*}

The proposed framework is illustrated in Fig. \ref{whole_networkstructure}. The input image, represented by $x$, has dimensions $W\times H\times 3$, where $W$ and $H$ are its width and height, respectively. The framework primarily consists of the core networks ($g_{a1}$, $g_{a2}$ and $g_{s}$) and the hyper networks ($h_{a}$ and $h_{s}$).

The two core encoder networks, labeled as $g_{a1}$ and $g_{a2}$, are tasked with learning two compact latent representations $y_{1}$ and $y_{2}$ from the input image. The architectures of  $g_{a1}$ and $g_{a2}$ both integrate two simplified attention modules, three residual group blocks (highlighted in cyan in Fig. \ref{whole_networkstructure}), and four stages of pooling operations. The residual group blocks are composed of four basic residual blocks \cite{resblock} connected in series.

We employ two branches to capture different resolutions of the original image. The first branch learns the high-resolution latent representation $y_{1}$, utilizing a $3 \times 3$ convolution as its downsampling module. In contrast, the second branch captures the low-resolution latent representation $y_{2}$ of the input image, leveraging a $1 \times 1$ convolution for downsampling.

To enable parallel entropy decoding of the quantized latents $y_{1}$ and $y_{2}$, we do not use any serial context model. As in \cite{channel, Liu_2023_CVPR}, we use channel-wise entropy model to encode and decode  $y_{1}$ and $y_{2}$ in parallel. However, it's noted that we first encode and decode $y_{1}$ in parallel, and then use $y_{1}$ as conditional side information to encode and decode  $y_{2}$ in parallel. We will provide a detailed explanation of this encoding and decoding process in Sec. \ref{conditioned_conding} and illustrate it in Fig. \ref{channel_wise_entropy_model}.

Subsequently, arithmetic coding compresses $\hat{y_{1}}$ and $\hat{y_{2}}$ into a bitstream. The decoded values, $\hat{y_{1}}$ and $\hat{y_{2}}$, are concatenated and forwarded to the primary decoder network $g_{s}$. This decoder mirrors the core encoder network $g_{a}$, with convolutions replaced by deconvolutions. While most convolution layers employ the leaky ReLU activation function, the final layer in both the hyperprior encoder and decoder operates without any activation function

\subsection{Dual-Branch Main Encoder Networks}
\label{sec_msrb}

We use two separate encoding networks to learn different resolution latent representations of the input image, and these two branch encoding networks do not share any information. However, it is important to note that the sizes of the downsampling convolution kernels we use are different, ensuring that they can capture distinct information from the input image. Larger convolution kernels are capable of learning global information from the input image, while smaller convolutions can capture local information, making it easier to reduce spatial redundancy in the data.

\subsection{Channel-Wise Auto-Regressive Entropy Model Based on Conditional Information Coding}
\label{conditioned_conding}

We use two encoding networks to learn latent representations of the input image at different resolutions, denoted as $y_{1}$ and $y_{2}$. The first encoding sub-network utilizes larger convolution kernels to capture the global information from the input image, preserving information that is more global in nature in the latent representation. The second encoding sub-network uses smaller convolution kernels to focus on the local information of the input image.

\begin{figure*}[!thp]
	\centering
		\includegraphics[scale=0.5]{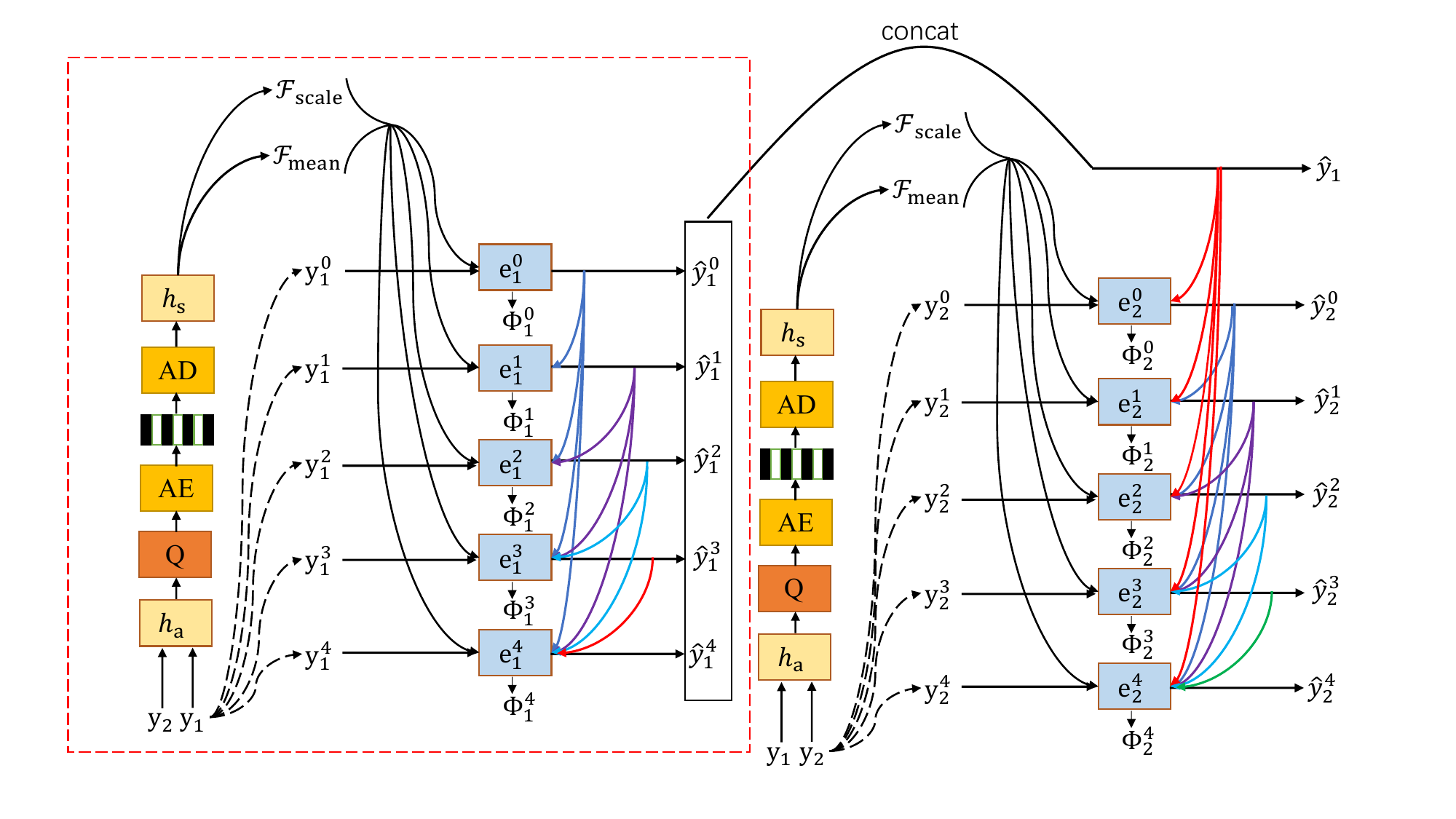}
	\caption{The two-stage channel-wise auto-regressive entropy model.The network of the $\phi$ have the similar networks where we just remove swin-transformer-based attention (SWAtten) modules from  $\phi$. }
	\label{channel_wise_entropy_model}
\end{figure*}

We use $y_{1}$ as auxiliary information to provide side information to $y_{2}$, thereby enhancing the efficiency of encoding and decoding for $y_{2}$. Since $y_{1}$ can provide $y_{2}$ with global information, it helps to eliminate redundancy in $y_{2}$.

As in \cite{channel, Liu_2023_CVPR},  we also use the channel-wise
auto-regressive entropy model to encode and decode $y_{1}$ and $y_{2}$. The detailed processing is shown in Fig. \ref{channel_wise_entropy_model}. As in \cite{Liu_2023_CVPR},  We evenly divide the channels of $y_{1}$ and $y_{2}$ into five slices. The channel number of $y_{1}$ and $y_{2}$ are fixed at 320, so each slice has 64 channels. Our channels are encoded and decoded in sequence, where later channels can fully utilize the information from preceding channels as prior knowledge, thereby reducing spatial redundancy between channels. When encoding and decoding the information of $y_{2}$, we can incorporate the information from $y_{1}$ into the encoding process of $y_{2}$. Every encoded or decoded slice can make use of $y_{1}$ as conditional side information.

During encoding and training, we can obtain the $\hat{y}_{1}$ and $\hat{y}_{2}$ in parallel. Since the values of $\hat{y}$ are available during these phases, encoding and training of $\hat{y}_{1}$ and $\hat{y}_{2}$ can proceed concurrently.

However, during decoding, as we cannot access all values of latent representations $\hat{y}_{1}$ and $\hat{y}_{2}$ simultaneously, we must decode them sequentially. Given that a subsequent slice relies on information from the preceding slice, these slices are decoded in sequence. However, the individual elements within each slice can be decoded in parallel.

Finally, we can combine $\hat{y}_{1}$ and $\hat{y}_{2}$ to obtain the decoded $\hat{y}$.

\subsection{Training}
The training images are obtained from the CLIC dataset ~\footnote{\url{http://www.compression.cc/}} and the LIU4K dataset \cite{LIU_dataset}. These images are randomly resized to a resolution of $2000 \times 2000$. Through data augmentation methods, such as rotation and scaling, we generate a collection of 81,650 training images, each with a resolution of $384 \times 384$.

Our proposed models are optimized using two distortion metrics: mean squared error (MSE) and multi-scale structural similarity (MS-SSIM). For the MSE-optimized, we choose $\lambda_{1}$ values from the set ${0.0016,0.0032,0.0075,0.015, 0.03, 0.045, 0.06}$. Each selected 
$\lambda$ initiates the training of a distinct model tailored for a particular bit rate. The filter number $N$ for latent representation is set at 128 for all bit rates. For MS-SSIM, $\lambda$ sequentially takes on values 12, 40, 80, and 120. All $\lambda$ values $N$ remains 128, and the channel numbers $M$ is set to 320 in the latent representation. Each model undergoes $1.5 \times 10^{6}$ training iterations using the Adam optimizer with a batch size of 8. The starting learning rate is set at $1 \times 10^{-4}$ for the first 750,000 iterations, then it gets halved after every subsequent 100,000 iterations. 

\section{Experimental Results}
\label{sec_results}

In this section, we compare some recent learned image compression methods and traditional image codecs with our proposed method  in terms of Peak Signal-to-Noise Ratio (PSNR) and MS-SSIM  metrics. 
The performance of different schemes are evaluated in two datasets with different resolutions. The  Kodak PhotoCD dataset \footnote{\url{http://r0k.us/graphics/kodak/}} is comprised  of 24 images with a resolution of $768 \times 512$.  Some recent learning-based schemes includes GLLMM \cite{GLLMM}, He2021 \cite{He_2021_CVPR}, Hu2021 \cite{Hu_AAAI}, and Cheng2020 \cite{Cheng2020_paper}. The traditional image codecs includes the latest image codec VVC-Intra (4:4:4) \footnote{\url{https://vcgit.hhi.fraunhofer.de/jvet/VVCSoftware_VTM/tree/VTM-5.2}}, BPG-Intra (4:4:4), JPEG2000, and JPEG.

For a fair comparison, we have implemented the method described in Cheng2020 \cite{Cheng2020_paper}, increasing its number of filters $N$ from 192 to 256 at high rates. This modification leads to enhanced performance compared to the original results in \cite{Cheng2020_paper}. The results from He2021 \cite{He_2021_CVPR} are sourced from the code available at \cite{lu2022high}.

\subsection{Rate-Distortion Performances}

\begin{figure*}[t]
\centering
\subfigure
{\begin{minipage}[t]{0.5\linewidth}
\centering
\includegraphics[scale=0.45]{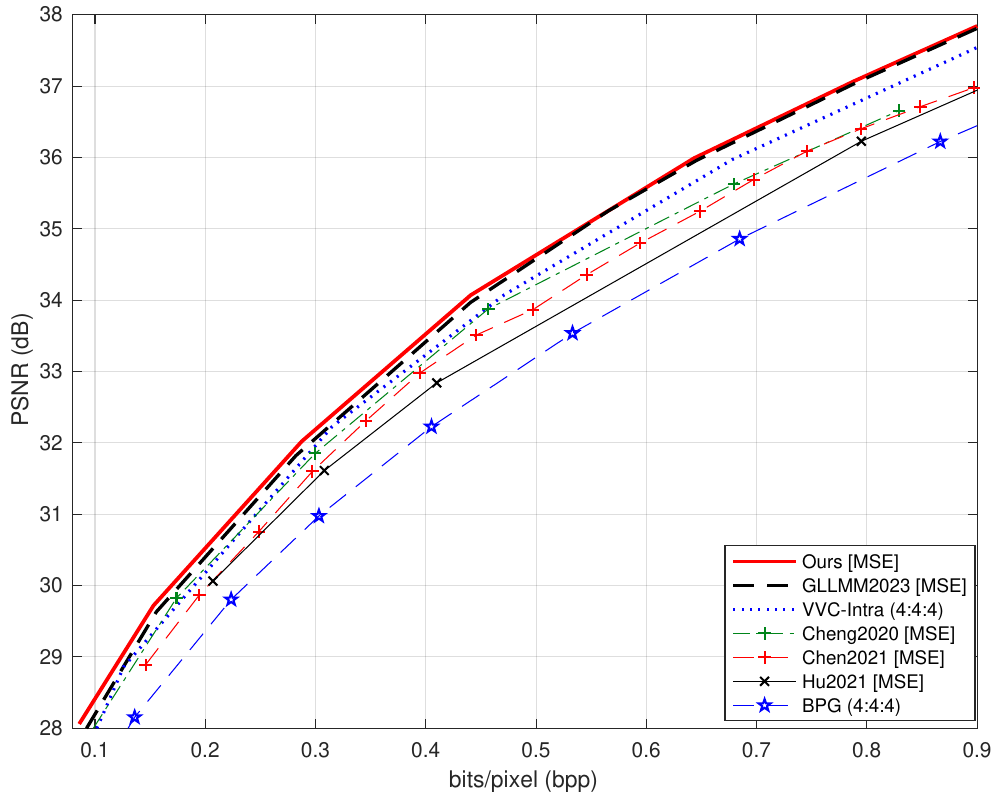}
\end{minipage}
}%
\subfigure
{\begin{minipage}[t]{0.5\linewidth}
\centering
\includegraphics[scale=0.45]{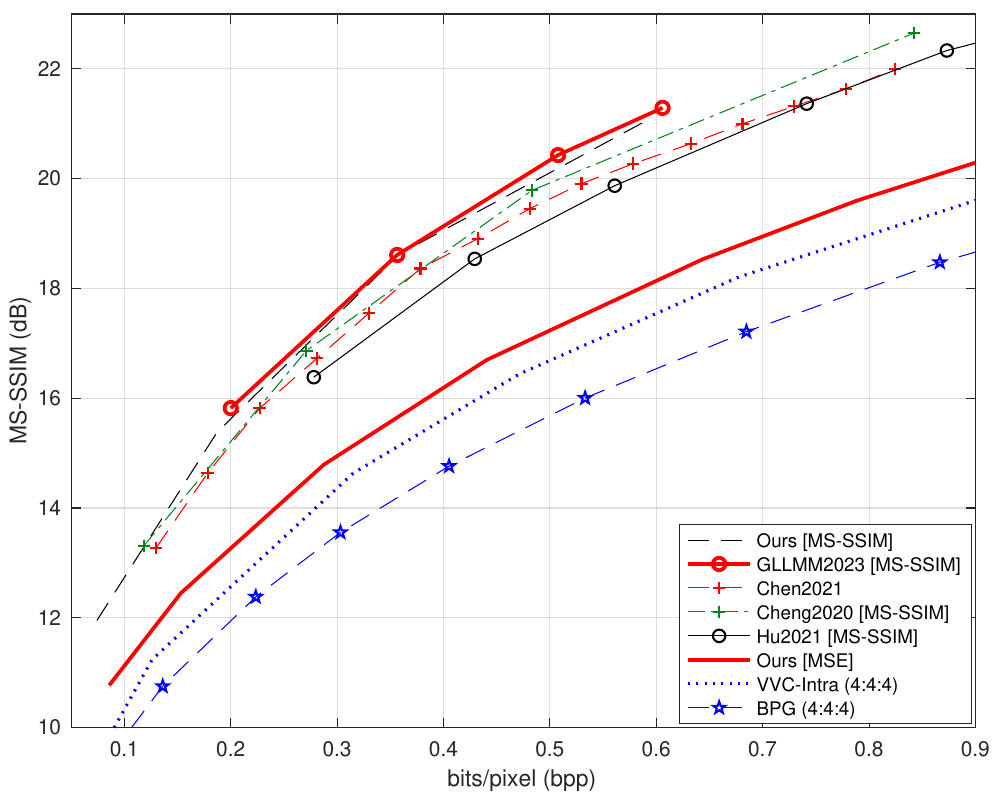}
\end{minipage}
}%
\centering
\caption{The average PSNR and MS-SSIM performance of the 24 images in the Kodak dataset.}
\label{fig:kodak}
\end{figure*}

Fig. \ref{fig:kodak} depicts the average R-D curves of various methods evaluated on the Kodak dataset. Among the PSNR-optimized methods, GLLMM (MSE) \cite{GLLMM} achieves the best performance in other methods, surpassing even VVC (4:4:4). Our method closely matches the coding performance of GLLMM at high bit rates and achieves better performance at low bit rates.  Our method has a 0.3-0.35 dB gain over VVC (4:4:4) at all bit rates. Regarding MS-SSIM, our method slightly outperforms GLLMM. 

\subsection{Complexity and Performance Trade-off}

\begin{table*}[!t]
\caption{Comparisons of encoding and decoding time, BD-Rate saving over VVC, and model sizes of the low bit rates and high bit rates.}
\begin{center}
\begin{tabular}{|c|c|c|c|c|c|c|}
\hline
\textbf{}& \textbf{Method} & \textbf{Enc.} & \textbf{Dec. } & \textbf{BD-Rate} &\textbf{Model(L) }&\textbf{Model(H)} \\ 
\hline
\multirow{6}{*}{Kodak}  & VVC  & 402.3s& 0.61s& 0.0 & 7.2 MB & 7.2MB\\ \cline{2-7} 
                         & Hu2021 \cite{Hu_2021}  &35.7s& 77.3s& 11.1 \% & 84.6 MB & 290.9MB\\ \cline{2-7} 
                         & Cheng2020 \cite{cheng2020}  &26.4s& 28.5s& 2.6 \% & 50.8 MB & 175.2MB\\
                         \cline{2-7} 
                         & He2021 \cite{He_2021_CVPR}  &24.4s& 5.2s& 8.9 \% & 46.6 MB & 156.6 MB\\
                         \cline{2-7} 
                         & GLLMM \cite{GLLMM}  &467.9s& 467.9s& -3.13\% & 77.1 MB & 241.0MB\\ \cline{2-7} 
                         & \textbf{Ours}  &\textbf{2.2 s}& \textbf{2.5s}& \textbf{-4.35\%} & \textbf{68.6 MB} &  \textbf{68.6MB}\\ \cline{2-7}
                         \hline
\end{tabular}

\label{runing_time}
\end{center}
\end{table*}

\begin{table*}[!t]
\caption{Comparisons of encoding and decoding time, BD-Rate saving over VVC, and model sizes of the low bit rates and high bit rates.}
\begin{center}
\begin{tabular}{|c|c|c|c|c|c|c|}
\hline
\textbf{}& \textbf{Method} & \textbf{Enc.} & \textbf{Dec. } & \textbf{BD-Rate} &\textbf{Model(L) }&\textbf{Model(H)} \\ 
\hline
\multirow{6}{*}{Kodak}  & VVC [4] & 402.3s& 0.61s& 0.0 & 7.2 MB & 7.2MB\\ \cline{2-7} 
                         & Hu2021 [5]  &35.7s& 77.3s& 11.1 \% & 84.6 MB & 290.9MB\\ \cline{2-7} 
                         & Cheng2020 [1]  &26.4s& 28.5s& 2.6 \% & 50.8 MB & 175.2MB\\
                         \cline{2-7} 
                         & He2021 [6]  &24.4s& 5.2s& 8.9 \% & 46.6 MB & 156.6 MB\\
                         \cline{2-7} 
                         & GLLMM [2]  &467.9s& 467.9s& -3.13\% & 77.1 MB & 241.0MB\\ \cline{2-7} 
                         & \textbf{Ours}  &\textbf{2.2 s}& \textbf{2.5s}& \textbf{-4.35\%} & \textbf{68.6 MB} &  \textbf{68.6MB}\\ \cline{2-7}
                         \hline
\end{tabular}

\label{runing_time}
\end{center}
\end{table*}

Table \ref{runing_time} illustrates a comparative results of average encoding/decoding times, BD-Rate savings relative to VVC \cite{BDRate}, and model sizes at both low and high bit rates across various methods. Due to the non-deterministic issues encountered in GLLMM \cite{GLLMM}, VVC, Hu2021 \cite{Hu_AAAI}, and Cheng2020 \cite{cheng2020} when executed on GPU, we conducted evaluations exclusively on a common CPU platform, namely the 2.9GHz Intel Xeon Gold 6226R CPU, to ensure fairness in the comparisons.

Compared to the GLLMM \cite{GLLMM}, our method is much faster in encoding and decoding, about 200 times quicker. Our model works better and is smaller in size.

Compared to Cheng2020 \cite{cheng2020}, our method is much faster in both encoding and decoding, being approximately 11 times quicker. Additionally, our rate-distortion performance outperforms that of Cheng2020 by 6.95\%. Compared to \cite{He_2021_CVPR}, not only is our encoding and decoding speed superior, but our R-D performance also shows a significant improvement, outperforming it by roughly 13.25\%.

\subsection{Performance Improvement of Different Modules}
\label{Ablation}
\begin{table}[!thp]
\caption{The performance  of different modules}
\begin{center}
  \begin{tabular}{cccc}
  \hline
 \textbf{Module} & \textbf{Bit rate}  & \textbf{PSNR } & \textbf{MS-SSIM }\\
  \hline
  \textbf{Ours} & 0.1531 & 29.72 dB & 12.67 dB \\
  \textbf{w/o CI} & 0.1582 & 29.63 dB& 12.59 dB \\
  \textbf{w/o TB}& 0.1632 &29.51 dB &12.48 dB \\
  \hline
  \textbf{Ours} & 0.9013 & 37.85 dB & 20.53 dB \\
  \textbf{w/o CI} & 0.9123 & 35.78 dB&20.48 dB \\
  \textbf{w/o TB}& 0.9146 &35.62 dB &20.36 dB\\
  \hline
\end{tabular}
\label{Tab:abalation}
\end{center}
\end{table}
Table \ref{Tab:abalation} compares the results when we do use conditional information (CI) and two branches (TB) main encoder  respectively, and the other modules remain the same. It can be seen that the performance drops by about 0.1 dB without conditional information (CI). The performance drops by about 0.2 dB at both low and high bit rates without two branches (TB).

\subsection{Performance Comparison of Different Group Partitions}
\label{Ablation}
\begin{table}[!thp]
\caption{The performance of different groups}
\begin{center}
  \begin{tabular}{ccccc}
  \hline
 \textbf{Groups} & \textbf{Bit rate}  & \textbf{PSNR } & \textbf{MS-SSIM }& \textbf{model Size}\\
  \hline
  \textbf{5} & 0.1531  & 29.72 dB & 12.67 dB &68.6  MB \\
  \textbf{10} & 0.1548 & 29.75 dB &12.69 dB  &70.2 MB\\
  \hline
  \textbf{5} & 0.9013 & 37.85 dB & 20.53 dB &68.6  MB\\
  \textbf{10}& 0.9023  & 37.90 dB &20.57 dB &70.2 MB\\
  \hline
\end{tabular}
\label{Tab:abalation_spilt}
\end{center}
\end{table}

We also explore scenarios where the latent representations $y_{1}$ 
 and $y_{2}$	are evenly divided into five and ten groups, respectively, with the results shown in Table \ref{Tab:abalation_spilt}. It can be observed that when the latent representations are divided into ten groups, the performance increases only slightly. Additionally, the model size significantly increases. This is attributed to the fact that dividing the channels into too many groups results in fewer channels per group, making it challenging to reduce spatial redundancy between pixels.

\section{Conclusions}

In this paper, we propose two techniques aimed at improving coding performance and speeding up the decoding process. These techniques consist of a dual-branch encoder network and a conditional information module. We employ two independent encoding networks to learn the latent representations of input images at different resolutions, which facilitates the reduction of spatial redundancy in the input images. The high-resolution latent representation primarily captures the global information of the input image, while the low-resolution latent representation predominantly represents its local details. We utilize the high-resolution latent information as side information for the low-resolution latent representation. This approach aids in reducing the spatial redundancy of the low-resolution latent representation, thereby enhancing encoding efficiency. We also employ a parallel channel-wise auto-regressive entropy model \cite{channel, Liu_2023_CVPR} for encoding and decoding low-resolution and high-resolution latent representations.


\section{Acknowledgments}
\label{sec:Acknowledgments}

This work was supported by the National Natural Science Foundation of China (No. 61474093), Industrial Field Project - Key Industrial Innovation Chain (Group) of Shaanxi Province (2022ZDLGY06-02), the Natural Sciences and Engineering Research Council of Canada (RGPIN-2020-04525), Google Chrome University Research Program, NSERC Discovery Grant RGPIN-2019-04613, DGECR-2019-00120, Alliance Grant ALLRP-552042-2020; CFI John R. Evans Leaders Fund.


\end{document}